\documentclass{article}

\usepackage{arxiv}

\usepackage[utf8]{inputenc} 
\usepackage[T1]{fontenc}    
\usepackage{hyperref}       
\usepackage{url}            
\usepackage{booktabs}       
\usepackage{amsfonts}       
\usepackage{nicefrac}       
\usepackage{microtype}      
\usepackage{lipsum}
\usepackage[utf8]{inputenc}
\usepackage{graphicx}

\title{The Evolution of Decentralized ICT Networks}

\author{
  Edmond Shami\thanks{Edmond holds a Bsc in Telecommunications and Electronics Engineering from Jordan University of Science and Technology. He is also a Cisco CCNA and CCDA certified engineer and  has 3 years of experience in telecommunications design from Dar Al Handasah, where he is currently employed.} \\
  Telecom Design Engineer\\
  Dar Al-Handasah\\
  Amman, Jordan 11191 \\
  \texttt{edmond.shami@gmail.com} \\
   \
 \\
  \\
  \\
  \\
  \\
}

\begin{document}
\maketitle
\begin{abstract}
Traditional networking systems, especially the famous 3-tier topology design, focused more on centralizing the networking systems, and used redundancy as a protection mechanism against future failures of the system. However, in recent years, the evolution of decentralization has taken place through several design techniques. Starting with clients requiring physically separate networks for critical applications. And then emerged the Spine Leaf topology and the Software-Defined Network solutions. In this research paper, I will first, prove how decentralization is better than centralization using aspects of probability theory. Second, I’ll go further and show how the Spine-Leaf and Software-Defined Networks are inherently decentralized, redundant and are better than the traditional centralized networks.
\end{abstract}


\section{Introduction}
What was usually called Scaling-Up is now being replaced with Scaling-Out, not only in the Networking Systems, but also in the Storage Systems. Scaling-Up systems, after a certain threshold, adding more unites won't increase performance, unlike Scaling-Out, with almost no limits to growth capabilities. I also go on to explain the benefits of decentralizing systems compared to centralized systems, relying on inequalities in concave transformations and will go further in showing the effects of hidden harm in centralized systems. I will venture after that to show how decentralization is naturally taking place in ICT networking systems, by mainly focusing on the Spine-Leaf, its benefits compared to the traditional 3-tier topology and how it is paving the way to more decentralized systems such as the Software-Defined Networks and the Software-Defined Storage.

\section{Decentralization VS Centralization}

\subsection{How the effect of errors is less in decentralized systems than in centralized systems.}
If we assume that harm is unbounded and nonlinear function in the \(C^2\) classification, following this form:
\begin{equation}
H_(x)= -kx^\beta                                              
\end{equation}

Where x represents error and $k \in (0,\infty)$ and $\beta \in [0, \infty)$

\begin{figure}[htp]
    \centering
    \includegraphics[width=9cm]{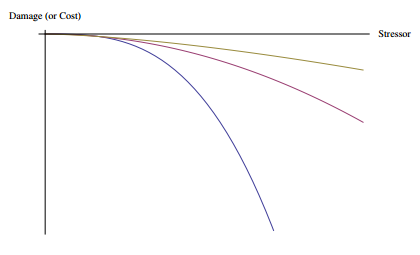}
    \caption{Harm Function: k = 1, $\beta =
3/2, 2, 3$. Source: The Technical Incerto Vol.2}
    \label{fig: H(x)}
\end{figure}


B is the unit subject to harm, so the resulting function is $\sigma_(x) = B + H_(x)$

If we further segment x into fragments, using $w_{i}$, where $0 \leq w_{i} \leq 1$, we can prove using inequalities in concave transformations that:
\begin{equation}
H_(\sum_{i=1}^{N} w_{i}x) \leq \sum_{i=1}^{N} H_(w_{i} x)
\end{equation}
Hence:

\begin{equation}
E_(\sigma_(B)) \leq E_(\sum_{i=1}^{N} \sigma_(w_{i}B))
\end{equation}

\subsubsection{The big effect of unseen harms in centralization.}

Assume that x (error) follows a simple Pareto distribution:

\begin{equation}
\rho_{\alpha, L}(x) = (\alpha L^\alpha)/ (x^{\alpha+1})  
\end{equation}

If we further segment x into N fragments, where $\xi_ = x/N$ and $N \geq 1$, the distribution of the resulting harm function will be $g_(\rho  \circ  h)(z)$:

\begin{equation}
g_{\alpha, L, N}(\xi) = (\alpha^\alpha N^{-\alpha} (-\xi/k)^{-\alpha/\beta})/(\beta \xi)
\end{equation}

The resulting mean of loss for $\xi < -k(L/N)^\beta$ and with $a > (1+\beta$):

\begin{equation}
M_{\beta}(N) = \int_{-\infty}^{-kL \beta /N} \xi g_{\alpha, L, N}(\xi) d\xi = - (\alpha kL^\beta N^{\alpha (1/\beta - 1)-1}) / (\alpha - \beta) 
\end{equation}

The ratio of K of N of its fragments compared to the original mean, results in the degradation in the mean, as we can see below:

\begin{equation}
K M_{\beta}(KN)/ M_{\beta}(N) = K^{\alpha (1/\beta - 1)}
\end{equation}

\begin{figure}[htp]
    \centering
    \includegraphics[width=9cm]{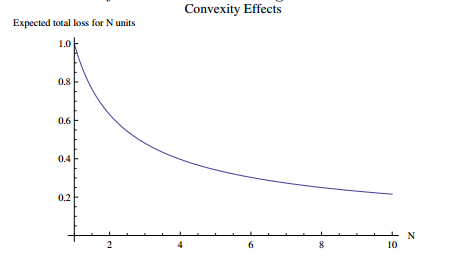}
    \caption{The mean harm in total as a result of concentration. Degradation of the mean for $N=1$ compared to a large N, with $\beta = 3/2$. Source: The Technical Incerto Vol.2}
    \label{fig: K Ratio}
\end{figure}

As we can see, the skewness indicates that for centralized systems, predicting the total effect of harm using large samples is not possible. Which indicates that we leave out unseen harm, which, even if small, can have a huge effect on the system, resulting in larger harm than expected. 

\section{Decentralizing the network topology.}

The most famous network design, was the 3 -ier architecture. Where the network consists of modular switches at the core and distribution layers, and expansion is done using scaling-up (filling empty slots of the same switch). 

\begin{figure}[htp]
    \centering
    \includegraphics[width=15cm]{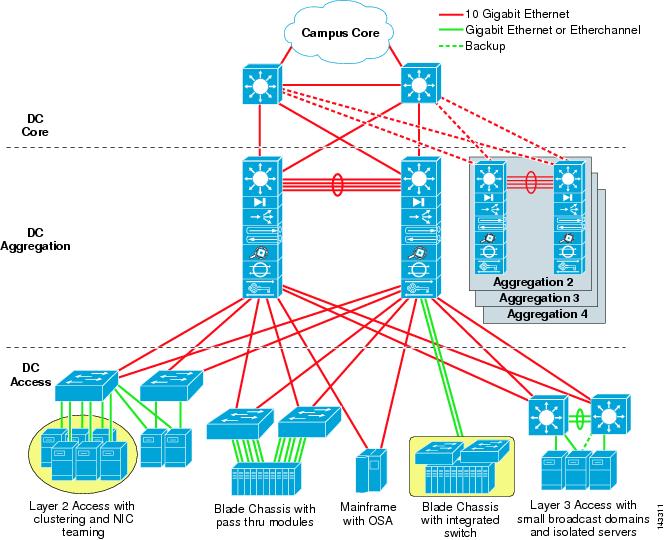}
    \caption{3-tier topology. Source: Cisco}
    \label{fig: 3-Tier}
\end{figure}


This model is well known for its centralized features, where the core controls the whole network including connectivity to the outside network. And its built-in redundancy includes connecting the core switches to each other for easier handover when one of them is down. Not to forget that this design introduces more hops and design complexity with more inter-connectivity between network elements, introducing the risk of silent packet drops which are commonplace in modular switches. Hence, less visibility into network traffic. That’s why, the Sigmoid function better represents the limits of the future growth in the 3-tier design architecture, because the growth is limited to the number of slots available in the modular chassis. The figure below shows a simple Sigmoid function where growth eventually terminates at one point in the future:

\begin{figure}[htp]
    \centering
    \includegraphics[width=8cm]{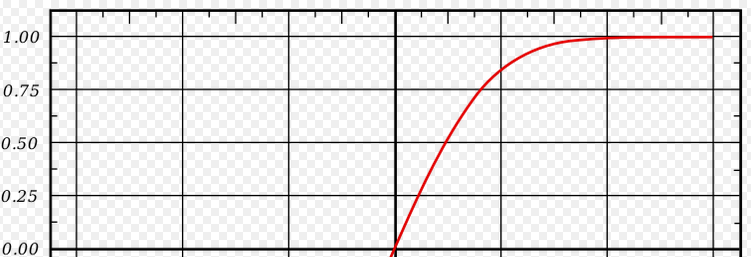}
    \caption{A simple Sigmoid function: $f_(x) = erf(x)$ $x \in R$, $F_(x) = 0$ at $x=0$ and $x \geq 0$. Where x represents switch modules.}
    \label{fig: Error Function}
\end{figure}

An additional flaw in the 3-tier architecture is that you can’t add more than 2 core switches. Which, moves us to the next point, which is, you have to predict the maximum demand of the network and its future growth from the early design stages, putting the network at more risk and harm by unpredictable/hidden future events and growth requirements.

\subsubsection{Early signs of adopting decentralization in design stages.}

Through my humble experience at Dar Al-Handasah, an engineering design firm, among the top 5 engineering design firms for buildings in the world, several clients (and in some cases country codes such as Qatar) require you to physically separate the network of critical systems (CCTV for example) from the rest of the ICT network, and in some cases,  the network components have to be in different rooms too. Such clients usually have higher security requirements, which included governments and airports. 

\subsubsection{The rise of Spine Leaf Design.}

The Spine Leaf introduced a solution to the design issues found in the classical 3-tier topology by:

\begin{enumerate}
\item Reducing the number of hops in the network (2 layers only)
\item Reducing the inter-connectivity of the network, because Spines don’t connect to each other and so are the Leaves. See Figure 5:
\end{enumerate}

\begin{figure}[htp]
    \centering
    \includegraphics[width=12cm]{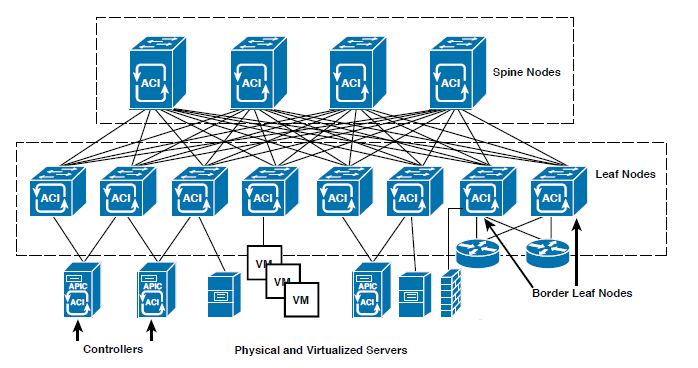}
    \caption{Spine-Leaf Architecture. Source: CCNA Data Center DCICT 200-155}
    \label{fig: Spine-Leaf}
\end{figure}

With these two features, Spine Leaf transferred the growth of networks from a Sigmoid like growth to an almost linear growth, where growth is almost unlimited. You simply add a Spine/leaf whenever the need requires. Also, this introduced more decentralization to the network and smaller fault domains. 

Spines have no other functionality but to pass data between Leaves. And each Leaf is a domain on its own. This design allows network functionalities to be distributed among Leaves, as shown in the figure above. There can be Data Center Leaves, Border Leaves, DMZ Leaves, SDN Leaves and Campus Leaves and with each Leaf one hop away from the other. 

This design enabled the move towards Software-Defined Networks. 

\subsubsection{Software-Defined Networks (SDN)}

SDN is an emerging technology in networking systems in which the Control Plane and the Data Plane are decoupled in networking devices, paving the way for the control plane to be a new device (called SDN controller). Through the SDN controller, you can define the network policy in simple terms, and the network devices can configure themselves automatically. 

There are several degrees of decoupling between the control and data plane. The one recommended in this paper is the “Declarative” approach. Where the controller only defines the policy of the network without doing any further processing or management of the network data, which is left entirely to the other network devices. This paved the way to the introduction of APIs (Application Programming Interfaces) in networking systems. Giving more control to the user to create his/her own applications to configure, automate and monitor their networks. 
 
\section{Comments}

Besides the benefit of avoiding unseen risks in centralization through decentralizing systems, a comparison study done by Mellanox between Fixed Port Switches (Spine-Leaf Architecture) and Modular Switches (3-Tier Architecture), indicated that Fixed Port Switches:

\begin{enumerate}
\item Enjoy 75\% lower price per port
\item Consume 75\% fewer watts per port 
\item Provide smaller fault domains, compared to a Chernobyl-like effect in Modular Switches where half of the network is down if one of the switches melts down.
\end{enumerate}

At first, the main excuse for centralization was efficiency, forgetting the hidden risks in such deployments. Now the age of decentralization is disrupting the way networks are designed, making networks more resilient against unpredictable risks.

We can also see that decentralization is also paving its way through storage systems. As we witness the rise of the Software-Defined Storage solutions. My own recommendation is to also see more physical separation, but how will that take place, tinkerers and network administrators shall probably lead to this revolution with proven design concepts. 


\bibliographystyle{unsrt}  


\end{document}